\documentclass{kapproc} 
\usepackage{psfig} 
\kluwerbib

\begin{document}

\articletitle[]{Influence of internal energy on the stability of
relativistic flows}

\author{Manuel Perucho$^1$, Michal Hanasz$^2$, 
Jos\'e Mar\'{\i}a Mart\'{\i}$^1$, H\'el\`ene Sol$^3$}
\affil{$^1$Departament d'Astronomia i Astrof\'{\i}sica. 
Universitat de Val\`encia. Burjassot (Spain)}
\affil{$^2$Nicolaus Copernicus University. Toru\'n (Poland)}
\affil{$^3$Observatoire de Paris/Meudon (France)}

\begin{abstract}
A set of simulations concerning the influence of internal energy on
the stability of relativistic jets is presented. Results show that 
perturbations saturate when the amplitude of the velocity perturbation 
approaches the speed of light limit. Also, contrary to
what predicted by linear stability theory, jets with higher specific 
internal energy appear to be more stable.
\end{abstract}

\section{Introduction}

  The production of collimated relativistic outflows is common among 
extragalactic radio sources. The question of why some of these sources
produce jets that propagate up to hundreds 
of kpcs along nine decades in distance scale (as, e.g., Cyg~A) whereas 
others (e.g., 3C31) decollimate and flare after few kpc, arises
naturally. To give an answer, the analysis of the nonlinear stability 
of relativistic flows against the growth of Kelvin-Helmholtz (KH) 
perturbations emerges as a powerful tool. 

  Focusing on the relativistic regime, the linear growth of KH 
perturbations both in the vortex sheet approximation (see Birkinshaw
1991a for a review), and sheared flows (Birkinshaw  1991b, Hanasz \&
Sol 1996) has been extensively analyzed. However, the nonlinear regime 
remains practically unexplored with only a couple of papers (i.e., 
Rosen et al. 1999, Hardee et al. 2001) covering partial aspects of the 
problem. In the present work, we concentrate on the influence of the specific 
internal energy in the long term stability of relativistic flows. The 
reason for this relies on two points: i) it is a genuine relativistic
effect and no numerical study has been yet performed for it 
(Rosen et al. 1999 have reported preliminary results), 
and ii) a remarkable stability of extremely {\it hot} jets has been 
found in simulations (see, e.g., Mart\'{\i} et al. 1997), in 
contradiction with predictions of linear stability analysis.  
 
\section{Numerical simulations}
 
\begin{figure} 
\centerline{
\psfig{file=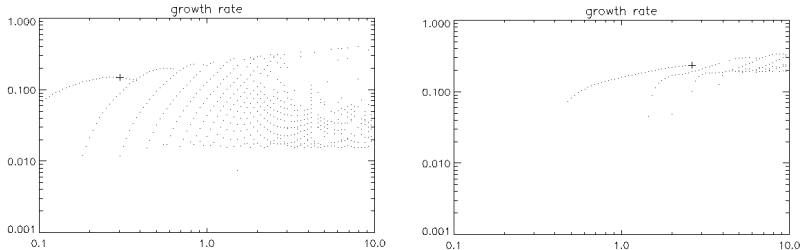,width=11cm,angle=0,clip=}  
}
\caption{Dispersion relation solutions for models A (left) and D
(right) in Table \ref{tab:param}. Panels show growth 
rates in terms of longitudinal wavenumber. Crosses indicate the
excited modes.} 
\label{fig:f1}
\end{figure}

  We focus on the simplest geometrical configuration of two dimensional 
planar relativistic flows and apply the temporal stability analysis,
i.e., we consider perturbations with real wavenumber and complex 
frequency, what gives modes which grow in time. For any equilibrium
model, the dispersion relation (for symmetric
modes in a slab relativistic jet; see, e.g., Hanasz \& Sol 1996) is
solved numerically for different modes (fundamental, body), getting 
solutions for real (proper frequency) and imaginary (growth rate) part 
of frequency as a function of wavenumber. The next step is to choose
the mode to excite. In our case, we choose the first body mode with
the highest growth rate (see Fig.~\ref{fig:f1}). 

  We start by generating steady two-dimensional slab jet models. A thin
shear layer between the ambient medium and the jet 
is used to achieve equilibrium. Due to symmetry
properties, only half of the jet has to be computed. Reflecting boundary
conditions are imposed on the symmetry plane of the flow, whereas
periodical conditions are settled on both upstream and downstream
boundaries. The steady model is then perturbed according to the 
selected mode, with an amplitude
$10^{-5}$ the background values. The mode wavelength is used as 
the grid size in the flow direction. The setup of the simulations
is given in Table~\ref{tab:param}.

\begin{table}[h] 
\caption{Simulation parameters}
\centering
\begin{tabular*}{\textwidth}{@{\extracolsep{\fill}}ccccccc}
\sphline
\it Model & \it $P$ & 
\it $\epsilon_b$ & \it $\epsilon_a$ &
\it $\nu$ & \it $\lambda$ \\
\sphline
A & $2.55 \, 10^{-3}$ & $0.08$ & $7.65 \, 10^{-3}$ & $0.11$ & $20.8$ \\
B & $0.01$            & $0.42$ & $0.04$            & $0.14$ & $9.80$ \\
C & $0.2$             & $6.14$ & $0.61$            & $0.44$ & $3.15$ \\
D & $2.0$             & $60.0$ & $6.0$             & $0.87$ & $2.39$ \\
\sphline
\end{tabular*}
\begin{tablenotes}
Labels $a$ and $b$ refer to ambient medium and jet, respectively.
In all simulations, the density in the jet is $\rho_b=0.1$, Lorentz
factor $W_b=5.0$, 
adiabatic exponent $\Gamma_{b,a}=4/3$. In the table, $P$ is pressure, 
$\epsilon_{a,b}$ is specific internal energy, $\nu$ is the
relativistic density ratio ($\rho_b (1+\epsilon_b)/\rho_a (1+\epsilon_a)$), 
and $\lambda$ is the wavelength of the excited mode (first body
mode). Throughout the paper, physical quantities are expressed in
units of the ambient density, $\rho_a$, the speed of light, $c$, and the
beam radius, $R_b$.
\end{tablenotes}
\label{tab:param}   
\end{table}

\begin{figure} 
\centerline{
\psfig{file=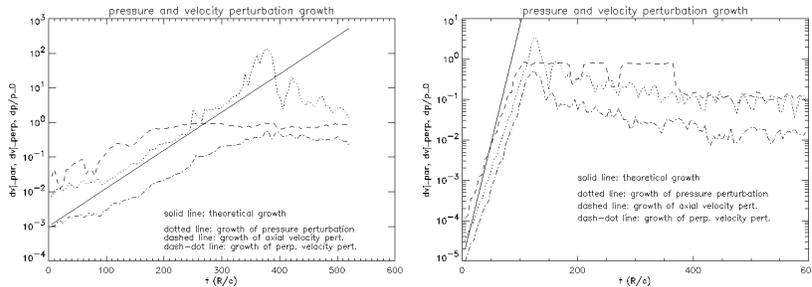,width=11cm,angle=0,clip=} 
}
\caption{Growth of pressure, axial and perpendicular velocity
  perturbation for models A (left) and D (right).}  
       
\label{fig:f2}
\end{figure}      

  The growth of the instability depends critically on the {\it
numerical viscosity} of the algorithm. Hence our first aim was to look 
for suitable numerical resolutions by comparing numerical and
analytical results for the linear regime. Resolution perpendicular to
the flow appeared to be essential requiring very high resolutions 
(400 zones/$R_b$) and thin shear layers with 40 to 45 zones. A (small)
resolution of 16 zones/$R_b$ along the jet was taken as a compromise
between accuracy and computational efficiency. Lower transversal 
resolutions and/or thicker shear layers led to non-satisfactory results,
with a slow or damped growth. Thinner
shear layers gave non-steady initial conditions.

\begin{figure} 
\centerline{
\psfig{file=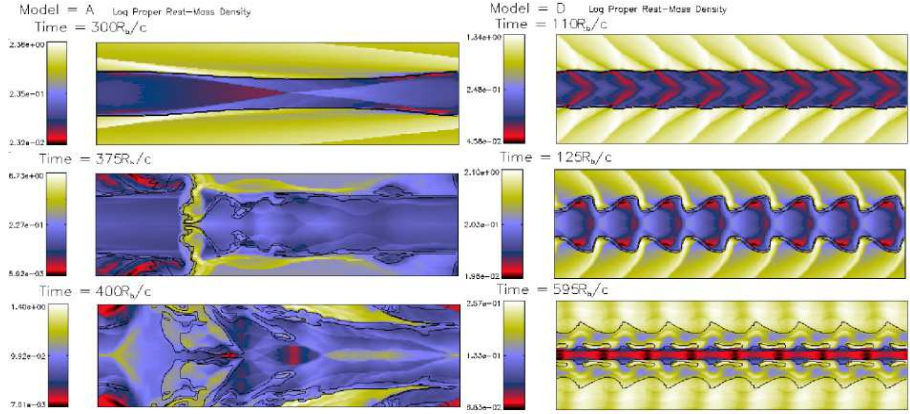,width=\textwidth,angle=0,clip=} 
}
\caption{Logarithm of rest mass density in three different instants
for models A (left) and D (right). From top to bottom, images
correspond to linear phase, mixing phase and disruption (model A), 
or last image of simulation (model D).}  
       
\label{fig:f3}
\end{figure}

  Two phases can be distinguished in all simulations: a phase of linear 
growth and a mixing phase, separated by the saturation point 
(Figs.~\ref{fig:f2}, \ref{fig:f3}). We checked that saturation is
reached when the velocity perturbation in the jet reference frame 
approaches the speed of light (Hanasz 1997). The mixing phase starts 
after saturation, 
in connection with the development of a planar shock within the jet 
flow and a wide shear layer. Disruption may occur besides the mixing
phase in some of the models. This phase is characterized by a complete
mixing of the ambient and jet materials and an effective spread of the
jet's axial momentum into the ambient. In the case of the simulations
presented in Table~\ref{tab:param}, disruption is perceived in models A and
B which, according to the linear stability analysis, have smaller
growth rates than models C and D.

\section{Conclusions}

  Saturation of the growth of perturbations is a very important
result (predicted on the basis of post-linear stability analysis by 
Hanasz 1997). After saturation, disruption is almost immediate, 
in the cases it occurs. Our results suggest that extremely hot jets
are more stable on the long term, contrary to the predictions of the
KH linear stability theory. Plots suggest that disruption is related to the 
pressure amplitude in the saturation point with respect to equilibrium 
value. It varies from a few in models C and D, to more than two orders 
of magnitude in models A and B (Fig.~\ref{fig:f2}). The fact that 
disruption occurs just after saturation suggests that the final fate
of the jet must be encoded in the jet properties at the end of the
linear phase, however we still have not found such a relation.

\end{document}